\documentclass[aps,prl,twocolumn,groupedaddress,showpacs,showkeys,floatfix]{revtex4-1}

\usepackage{amsmath,amssymb}
\usepackage[pdftex]{graphicx}
\usepackage{ascmac}
\usepackage{amsthm}
\usepackage{comment}
\usepackage{braket}
\usepackage{mathrsfs}
\usepackage{makeidx}

\usepackage{txfonts}

\newcommand{\be}{\begin{equation}}
\newcommand{\ee}{\end{equation}}

\begin{document}

\title{Topological Properties of Ultracold Bosons in One-Dimensional Quasiperiodic Optical Lattice}


\author{Fuyuki Matsuda}
\email[]{fuyuki.matsuda@scphys.kyoto-u.ac.jp}
\affiliation{Department of Physics, Kyoto University, Kyoto 606-8502, Japan}

\author{Masaki Tezuka}
\email[]{tezuka@scphys.kyoto-u.ac.jp}
\affiliation{Department of Physics, Kyoto University, Kyoto 606-8502, Japan}

\author{Norio Kawakami}
\email[]{norio@scphys.kyoto-u.ac.jp}
\affiliation{Department of Physics, Kyoto University, Kyoto 606-8502, Japan}

\date{\today}

\begin{abstract}
We analyze topological properties of the one-dimensional Bose-Hubbard model with a quasiperiodic superlattice potential. This system can be realized in interacting ultracold bosons in optical lattice in the presence of an incommensurate superlattice potential. We first analyze the quasiperiodic superlattice made by the cosine function, which we call Harper-like Bose-Hubbard model. We compute the Chern number and observe a gap-closing behavior as the interaction strength $U$ is changed. Also, we discuss the bulk-edge correspondence in our system. 
Furthermore, we explore the phase diagram as a function of $U$ and a continuous deformation parameter $\beta$ between the Harper-like model and another important quasiperiodic lattice, the Fibonacci model.
We numerically confirm that the incommensurate charge density wave (ICDW) phase is topologically non-trivial and it is topologically equivalent in the whole ICDW region.
\end{abstract}


\maketitle

\paragraph{Introduction.---}
Various materials forming quasicrystals have been studied for their unique physical properties.\cite{Shechtman,Fujiwara-Ishii,Bindi,Foerster,Wasio}
In recent years, topological phases have been a focus of constant attention in condensed matter physics.\cite{HasanKane, QiZhang} A typical example of topological phase is the integer quantum Hall effect, where quantized Hall conductances are proportional to the Chern numbers.\cite{vonKlitzing} The one-dimensional (1D) version of a quasicrystal is a quasiperiodic lattice. The relation between one-dimensional quasiperiodic systems and two-dimensional topological insulators has recently been pointed out and experimentally confirmed by using optical waveguides.\cite{Kraus} This idea provides a new point of view for both quasicrystals and topological phases. Accordingly, detailed analysis for topological phases in quasicrystals is needed. 

1D quasiperiodic models have theoretically been studied in various contexts.\cite{Harper, Azbel, AubryAndre, TKNN}
1D quasiperiodic systems can be realized not only by using optical waveguides, but also by using ultracold atoms.\cite{Fallani,Roati} An important difference between optical waveguide systems and ultracold atom systems is the existence of tunable particle-particle interactions. In the optical waveguide systems, photons do not interact with each other, but in the ultracold atom systems, it is possible to introduce and control the strength of interaction by using Feshbach resonance.\cite{Tanzi} Such advances have stimulated theoretical studies of 1D quasiperiodic systems with interaction.\cite{Roux,Adhikari,Schmitt,Orso,Albert,Cheng,Cai,Cetoli,Roscilde,QuasiperiodicFermi,Dhar,Dufour,Cai2013,Ribeiro}
Kraus \textit{et al.} showed that two different types of 1D quasicrystals can be smoothly connected without gap closing, which implies their topological equivalence.\cite{Kraus2}

In this Letter, motivated by the theoretical and experimental advances described above, we consider a Bose-Hubbard model with quasiperiodic modulation and analyze their topological properties.
We define the model hamiltonian, which gives the general definition of 1D boson quasiperiodic model with on-site interaction.
Then we introduce our methods for numerical analysis of the model. We also give our definition of the topological number 
in 1D quasiperiodic models.
After that,
first the basic topological properties of the Harper-type Bose-Hubbard model are shown. We discuss what happens as the interaction strength $U$ or the approximation accuracy changes.
Finally, we investigate whether a topological equivalence between Harper-type and Fibonacci-type models exists even in the interacting system. The phase diagram of the Bose-Hubbard model against the parameter characterizing the quasiperiodic potential  $\beta$ and interaction strength $U$ is obtained and region where topological equivalence holds is identified.
A summary of our findings concludes this Letter.

\paragraph{Model Hamiltonian.---}

We investigate the topological properties of the Bose-Hubbard model with quasiperiodic modulation described by the Hamiltonian defined on an $L$-site chain,
\be
\begin{split}
\hat{{\cal H}}(\phi, \beta) = &-t \sum_{\langle j, j' \rangle} \left( \hat{b}_{j'}^{\dagger} \hat{b}_{j} + \mathrm{H.c.} \right) \\ &+ \sum_{j}\bigg[\lambda V\left( \phi, j \right) \hat{n}_j + \frac{U}{2} \hat{n}_j (\hat{n}_j + 1) \bigg],
\end{split}
\label{harperfibonacciham1}
\ee
in which $\hat{b}_j^{\dag}(\hat{b}_j)$ is the boson creation (annihilation) operator, $\hat{n}_j = \hat{b}_j^{\dag} \hat{b}_j$ is the particle number operator, $t$ is the hopping strength, $\lambda$ is the modulation strength of quasiperiodic potential, $\phi$ is the phase of the modulation, $U$ is the on-site interaction strength, $V\left( \phi, j \right)$ are periodic functions of $j$ with period $(2\pi b)^{-1}$.
The hopping is introduced between neighboring sites $(j' = j + 1, j = 0,\ldots,(L-2))$, and also between the end sites $((j,j')=(L-1,0))$ with a phase factor $e^{i\theta}$ for the $\hat{b}_{1}^{\dagger} \hat{b}_L$ term for the twisted boundary condition (see below). In the following we take $t=1$ as the unit of energy.

\paragraph{Methods.---}

We apply the exact diagonalization and the density matrix renormalization group (DMRG) method to numerically obtain the many-body ground state of eq.~\eqref{harperfibonacciham1} for a given number $N$ of bosons. In order to investigate the topological properties of this system, we analyze the effect of boson number change in the open boundary condition and also calculate the Chern number by introducing twisted boundary conditions. The Chern number (with respect to $(\theta, \phi)$) is defined as 
\be
C = \frac{1}{2\pi i} \int_0^{2\pi} d\theta \int_0^{2\pi} d\phi \left[\Braket{\frac{d\Psi}{d\theta} | \frac{d\Psi}{d\phi}} - \Braket{\frac{d\Psi}{d\phi} | \frac{d\Psi}{d\theta}} \right]
\ee
in which $\theta$ is the phase of the twisted boundary condition. 
However, it is not possible to introduce twisted boundary conditions on a finite system with a genuine quasiperiodicity, because a quasiperiodic system does not have translation symmetry. Thus, we approximate the irrational number $b$, which characterizes the quasiperiodicity of the system, by a rational number so that we may impose twisted boundary conditions. We obtain the best rational approximations of the irrational number $b$ by using convergents of continued fraction representations. For example, for
\begin{equation}
b=\frac{3-\sqrt 5}{2} = 1-\cfrac{1}{1+\cfrac{1}{1+\dots}} \  , 
\end{equation}
the rational approximations will be 1/2, 1/3, 2/5, 3/8, 5/13, and so on. Additionally, we have used the method for calculating the Chern number from the ground states for a discrete set of values of $(\theta,\phi)$.\cite{FukuiHatsugaiSuzuki} In DMRG, we must use the same basis for four values of parameters forming a rectangle in the $(\theta,\phi)$-space to calculate link variables and lattice field strength which is defined in Ref.~\onlinecite{FukuiHatsugaiSuzuki}.

\paragraph{Topological Properties of the Harper-type Bose-Hubbard model.---}

\begin{figure}[t]
\includegraphics[width=8cm]{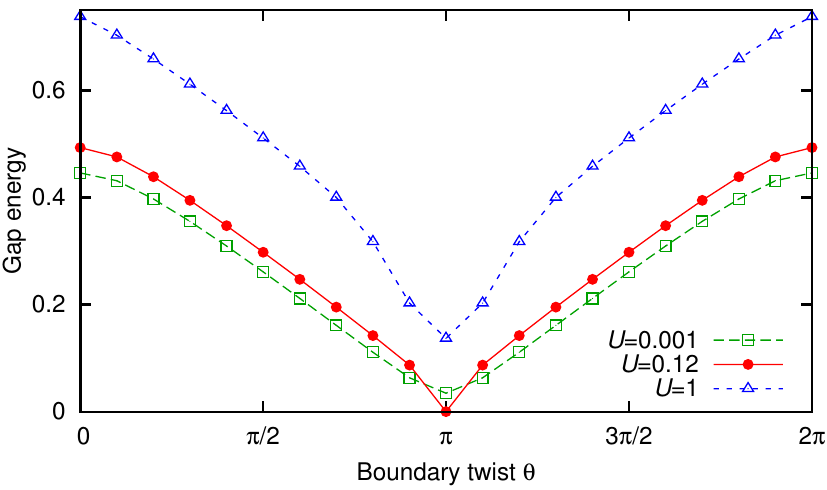}
\caption{Energy gap plotted as a function of $\theta$ in $[0, 2\pi]$, $\phi = 0$ plotted for three strengths of the interaction $U = 0.001, 0.12, 1.0$;
$\lambda = 1$, $(L, N) = (8, 3)$ for $b = 3/8$.}
\label{gapclose}
\end{figure}

1D Bose-Hubbard model can be boiled down to a one-particle problem of spinless fermions in the hard-core limit ($U=\infty$) by using a Jordan-Wigner transformation such as
\be
\hat{b}_j^{\dag}=\hat{c}_j^{\dag} \prod^{j-1}_{n=1} e^{i\pi \hat{c}_n^{\dag} \hat{c}_n}.
\ee
For the diagonal ($V^\mathrm{od}=0$) Harper-type modulation $V^\mathrm{d}\left( \phi, j \right) = \cos(2\pi b j + \phi)$, the new Hamiltonian is
\be
\hat{\mathcal{H}} = -t\sum_{\langle j j' \rangle}(\hat{c}_{j'}^{\dag} \hat{c}_j + \mathrm{H.c.}) + \lambda \sum_j \cos (2\pi b j + \phi) \hat{n}_j ,
\ee
in which $\hat{c}_j^{\dag}(\hat{c}_j)$ is the fermion creation (annihilation) operator and $\hat{n}_j = \hat{c}_j^{\dag} \hat{c}_j$.
When $b=p/q$ with coprime integers $p$ and $q$, which means that $b$ is rational, the system is periodic with a period $q$. In this condition, the energy band splits into $q$ bands. This band structure conforms with that of the electrons on square lattice under magnetic fields.
When $b$ is irrational, the band structure will be a cantor set and every band gap has a non-trivial Chern number.\cite{Hofstadter}

Now, we consider the case with a finite $U$. First, we approximate the quasiperiodic system by a periodic system (see (3)). When $U$ is sufficiently large, the system is similar to a free fermion system, and the gap energy will be close to the single particle level separation at the Fermi level in the fermion system. When $U$ is sufficiently small, almost all of the particles are in the one-particle ground state. Between these limits, we observe that the energy gap between the ground state and the first excited state closes for a certain choice of $(\theta,\phi)$ at a number of values of $U$. The gap closing behavior is shown in Fig.~\ref{gapclose} for an approximate system of $b = 3/8$. This result is consistent with the results by Deng and Santos, in which topological transitions are observed in the Bose-Hubbard model with a periodic superlattice.\cite{DengSantos} 

\begin{figure}[t]
\begin{center}
\includegraphics[width=8cm]{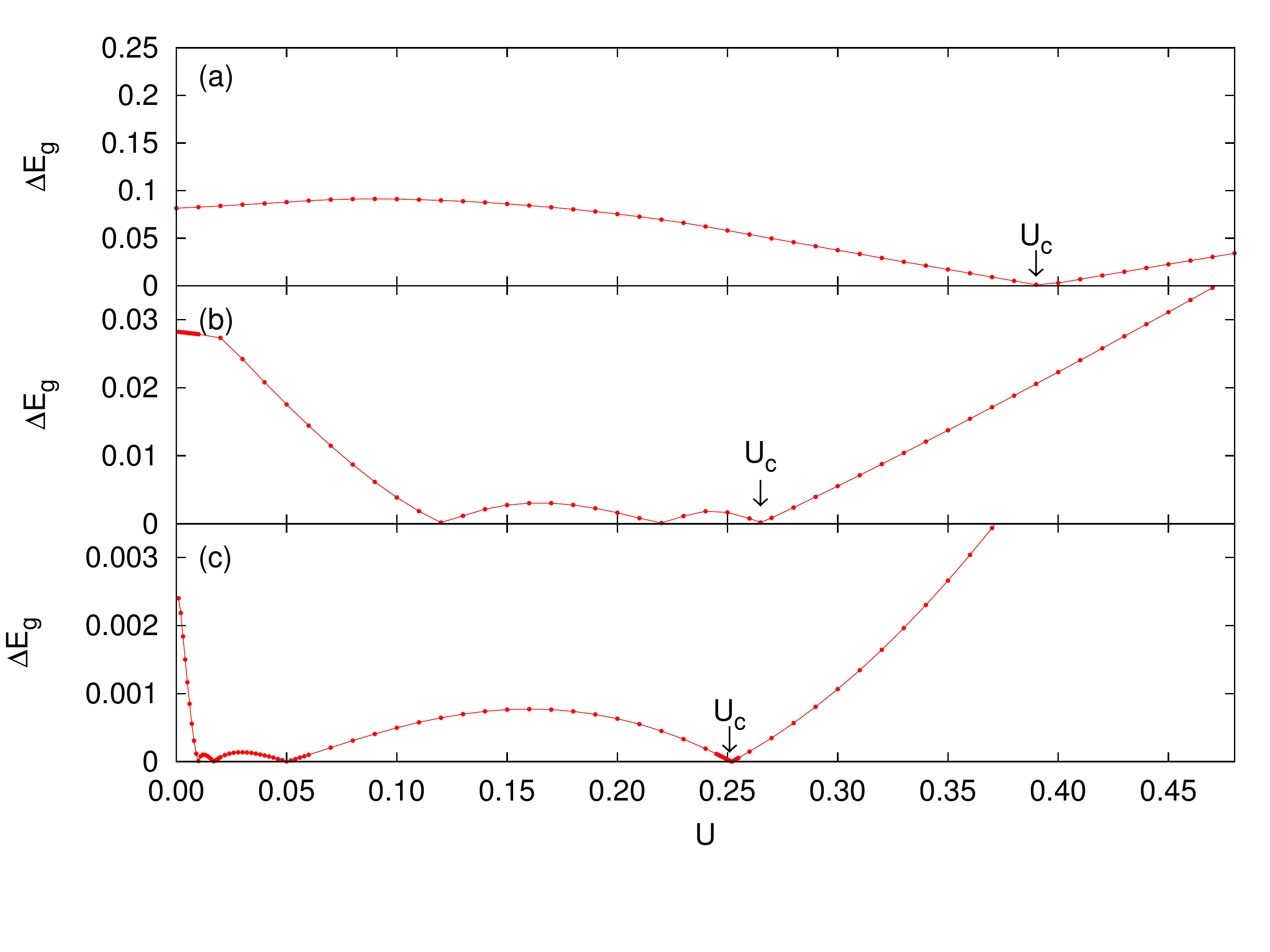}
\caption{The minimum energy gap $\Delta E_g$ in the $(\theta, \phi)$ space plotted as a function of $U$ for $t=1$, $\lambda = 1$. (a): $b=2/5$, $(L, N) = (5, 2)$. (b): $b=3/8$, $(L, N) = (8, 3)$. (c): $b=5/13$, $(L, N) = (13, 5)$. $U_c \simeq 0.39, 0.27, 0.25$ for (a), (b), and (c) respectively.}
\label{gap2}
\end{center}
\end{figure}

What happens as we improve the approximation of the quasiperiodicity in Eq.(3)? Fig.~\ref{gap2} shows $\Delta E_g$, which is the minimum of the energy gap between the ground state and the first excited state, in the parameter range $\phi \in [0, 2\pi), \theta \in [0, 2\pi)$ for $b = 2/5, 3/8, 5/13$. When the energy gap is zero for a certain choice of $(\theta,\phi)$, or $\Delta E_g$ becomes zero, the topological number is allowed to change its value. We define $U_c$ as the largest $U$ when the gap closes. We also calculate the Chern number. If $U<U_c$, the Chern number takes various values often changing by multiples of $L$. However, if $U>U_c$, for which the gap energy monotonically increases when $U$ increases. The Chern number remains $C=1$ in all accuracies $b=2/5, 3/8, 5/13$. This indicates that it is topologically non-trivial in the limit of quasiperiodic system ($b\rightarrow (3-\sqrt5)/2$).

\begin{figure}[t]
\centering \includegraphics[width=8.4cm]{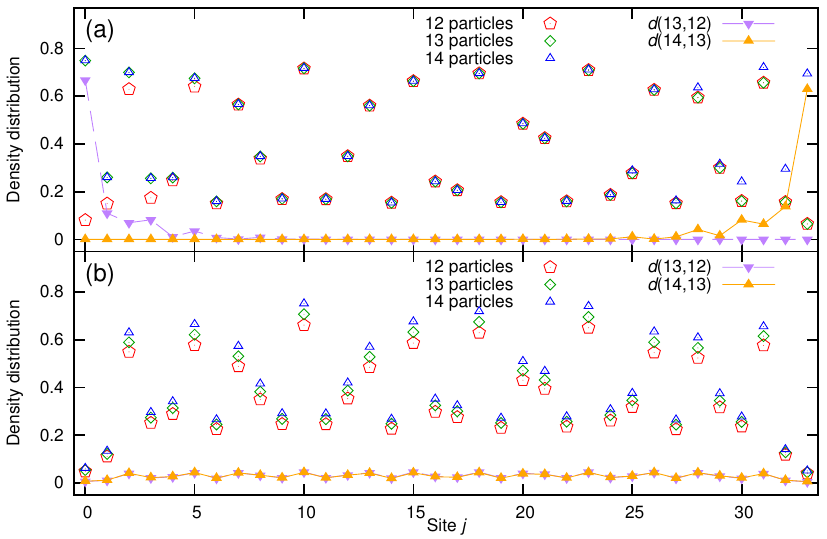}
\caption{The difference between the particle density distribution of the ground state of quasiperiodic Bose-Hubbard model with 34 sites, $b=\frac{\sqrt{5}-1}{2}$, $\phi=2$, $\lambda = 1$ and filling $\nu = 12/34, 13/34, 14/34$, (a) $U=100$ and (b) $1$. Red, green, blue points respectively show the particle density distribution $n_N(j)$ of $N$-particle systems for $N=12,13,14$, and purple and orange lines show the difference between them, $d(N+1,N)\equiv n_{N+1}(j) - n_N(j)$ for $N=12,13$.}
\label{U100_U1}
\end{figure}

In non-interacting fermionic systems, topologically protected edge states arise when the Chern number is non-zero and the system has an open boundary condition. This phenomenon is called the bulk-edge correspondence. In a 1D quasiperiodic system, such a topologically protected state arises as an end state, which is a localized state at the end of the system. In contrast, in interacting bosonic systems, we find that a non-trivial topological character appears in the particle density distribution. In our system, when the interaction strength $U$ is sufficiently large, such a localization can be seen in the difference in the density distribution between the ground states of different fillings, which is shown in Fig.~\ref{U100_U1}, calculated by DMRG in the open boundary condition.

The edge localization structure appears only in particular ranges of $\phi$. However, when $U$ is small, such an edge localization structure disappears. Despite the edge structure being absent in whole range of $\phi$, the Chern number remains non-zero. This implies that the edge localization structure does not have the same property as in the edge mode in non-interacting fermionic systems. Similar non-local density difference has been observed in the topologically non-trivial insulating phase in a superlattice Bose-Hubbard system.\cite{Grusdt}

\paragraph{Harper model and Fibonacci model.---}

Now we demonstrate that the ground states of the Harper-type and Fibonacci-type Bose-Hubbard models can be smoothly connected without closing the energy gap $\Delta E_g$ when the interaction is moderately strong. This implies that these two models are topologically equivalent. According to Ref.~\onlinecite{Kraus2}, we define a smooth modulation
\be
V(\phi, j, \beta)=\frac{\tanh \left[ \beta \left( \cos\left( 2\pi bj+\phi\right)-\cos\left(\pi b\right) \right) \right] }{\tanh \beta},
\ee
which becomes a Harper (Fibonacci) type one in the limit of $\beta \rightarrow 0$ ($\beta \rightarrow \infty$). Using this smooth modulation function, we consider the Hamiltonian as follows: 
\begin{align}
\hat{\cal H}(\phi, \beta) = &-t \sum_{\langle j j' \rangle} \left( \hat{b}_{j'}^{\dagger} \hat{b}_j + {\mathrm{H.c.}} \right) \nonumber\\
 &+ \sum_j\left[\lambda V\left( \phi+3\pi b, j, \beta \right)\hat{n}_j + \frac{U}{2} \hat{n}_j (\hat{n}_j + 1)\right].
\end{align}
\label{harperfibonacciham2}

\begin{figure}[t]
\begin{center}
\includegraphics[]{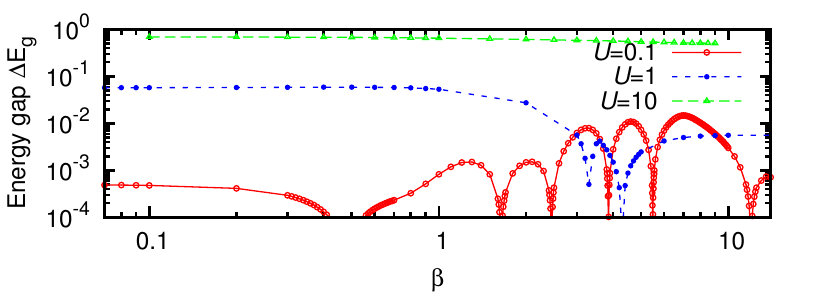}
\end{center}
\caption{The minimum energy gap between the ground state and the first excited state for the system of 13 sites with open boundary condition, $b=5/13$, $U=0.1$, $1$ and $10$, $\lambda=1$. When U is sufficiently large, while $\beta$ changes, the gap remains open and the topological number is fixed. This implies that Harper-like Bose-Hubbard model and Fibonacci-like Bose-Hubbard model are topologically equivalent in large $U$ region.}
\label{fig3}
\end{figure}

\noindent
We investigate the behavior of the energy gap $\Delta E_g$ while the parameters $\beta$ and $U$ changes. Figure \ref{fig3} shows the energy gap $\Delta E_g$ as a function of $\beta$.
Since $\Delta E_g$ does not become zero while $\beta$ changes for $U=10$, we conclude that the Harper-type and Fibonacci-type Bose-Hubbard models are topologically equivalent. The gap-closing behaviors for smaller $U$ are different. When $U=1$ and $U=0.1$,  $\Delta E_g$ becomes zero two and six times respectively. This behavior is complicated and sensitive to $U$. These gap-closing behaviors arise from finite size effects in the superfluid phase as discussed later.

\begin{figure}[t]
\begin{center}
\includegraphics[width=8.4cm]{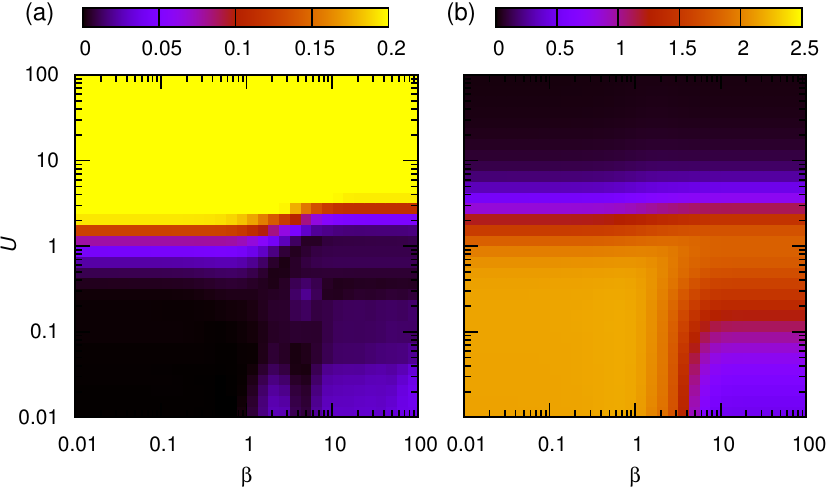}
\end{center}
\caption{(a) The minimum energy gap and (b) the superfluid density as a function of $U$ and $\beta$ computed for a system with (a) $b=5/13$, $(L, N) = (13, 5)$, (a) $b=8/21$, $(L, N) = (21, 8)$. Superfluid density is calculated for fixed phase $\phi =0$. }
\label{betaU}
\end{figure}

To investigate the phase diagram as a function of $\beta$ and $U$, we plot the energy gap $\Delta E_g$ and superfluid density $\rho_s$ in Fig. \ref{betaU}. The superfluid (SF) density can be computed using twisted boundary conditions:
\be
\rho_s = 2\pi L \frac{E_0^{\mathrm{apbc}}-E_0^{\mathrm{pbc}}}{\pi^2} ,
\ee
which is defined in Ref.\cite{Roux}, where $E_0^{\mathrm{pbc}}$ and $E_0^{\mathrm{apbc}}$ are the ground state energies for periodic ($\theta = 0$) and anti-periodic ($\theta = \pi$) boundary conditions. The SF density does not strongly depend on $\phi$. As $\beta$ is changed, $U_c$ changes its value. In the Harper-type region ($\beta \sim 0$), $U_c \sim 0.25$ in this approximation accuracy as we mentioned. However, in the Fibonacci-type region ($\beta \sim \infty$), $U_c \sim 1.55$. 

According to Ref.\cite{Roux}, in the Harper-type quasiperiodic Bose-Hubbard model, there are three phases in the quasiperiodic Bose-Hubbard model when the particle number density matches $b$, namely superfluid (SF), Bose glass (BG), and incommensurate charge density wave (ICDW). Figure \ref{betaU} (b) shows the superfluid density. This suggests that SF phase emerges at low $U$ and low $\beta$ region. At low $U$ and high $\beta$ region, the superfluid density is almost zero, but the energy gap is very small. These properties conform with the BG phase. The rest of the region ($U>U_c(\beta)$) is a gapful phase, which is ICDW. The above results support that this phase has a non-trivial Chern number $C=1$, and that the whole ICDW region is topologically equivalent.

\paragraph{Conclusion.---}

To summarize, we have considered a quasiperiodic Bose-Hubbard model with interaction by using exact diagonalization and DMRG.
First, we have shown that the energy gap of the Harper-type Bose-Hubbard model increases as the on-site interaction strength $U$ increases when $U>U_c$ ($\sim 0.25t$), and at the same time, topological number (Chern number) stays non-zero and constant. Also, while the topological equivalence between quasiperiodic models such as the Harper model and the Fibonacci model was only known in non-interacting fermionic systems previously, here we have shown that the topological equivalence exists also between interacting quasiperiodic systems, the Harper-type and Fibonacci-type Bose-Hubbard models. The equivalence exists when $U$ is moderately large, and the boundary of topologically non-trivial phase has been revealed. Moreover, the phase diagram as a function of $\beta$ and $U$ has been investigated. The phase diagram includes superfluid, Bose glass, and incommensurate charge density wave. 

\begin{acknowledgments}
NK thanks the Japan Society for the Promotion of Science for support through its FIRST Program and KAKENHI (Grants No. 22103005 and No. 25400366).
\end{acknowledgments}

\end{document}